\documentclass[twocolumn,prl,aps,superscriptaddress]{revtex4}
\usepackage{graphicx}
\usepackage{amsmath,amssymb}
\usepackage[latin1]{inputenc}
\begin{document}


\noindent\textbf{Comment on ``Intensity correlations and mesoscopic fluctuations
of diffusing photons in cold atoms''}

In a recent Letter~\cite{PRL}, O.~Assaf and E.~Akkermans claim that the
angular correlations of the light intensity scattered by a cloud of
cold atoms with internal degeneracy (Zeeman sublevels) of the ground
state overcome the usual Rayleigh law. More precisely, they found
that they become exponentially large with the size of the sample. In
what follows, we will explain why their results are wrong and, in
contrary, why the internal degeneracy leads to lower intensity
correlations.

Following the Authors' proposed experimental scheme and notations (their Eq.(3)), the
correlation of the transmission coefficient, averaged over the
positions and the internal states of the atoms, reads as follows:
\begin{equation}
\label{corr0}
\overline{T_{ab}T_{a'b'}}=
\overline{\sum_{ijkl}A_i^{\{R,m\}}A_j^{\{R,m\}*}A_k^{\{R,m'\}}A_l^{\{R,m'\}*}}
\end{equation}
As explained by the Authors, the configuration average of the
preceding equation leads to two possible pairing among the photon
paths, either $i=j, k=l$ which corresponds to
$\overline{T}_{ab}\overline{T}_{a'b'}$, or $i=l,
j=k$. In the latter case, 
the correlation of the transmission coefficient reads:
\begin{equation}
\label{corr1}
\overline{\delta T_{ab}\delta T_{a'b'}}=
\sum_{m,m'}\sum_{i}A^{\{m\}}_iA^{\{m'\}*}_i\sum_{j}A^{\{m'\}}_jA^{\{m\}*}_{j}.
\end{equation}
The preceding equation corresponds to Eq.(4) of the Letter,
with the only difference that the average over the internal states is
explicitly written (the configuration probabilities
$p(\{m\}),p(\{m'\})$ are left implicit). Since the sum over $i$ is
just the complex conjugate 
of the sum over $j$,  Eq.~\eqref{corr1} reads
\begin{equation}
\label{corr2}
\overline{\delta T_{ab}\delta T_{a'b'}}=
\sum_{m,m'}\left|\sum_iA^{\{m\}}_iA^{\{m\}'*}_i\right|^2.
\end{equation}
Now, for fixed ${\{m\},\{m'\}}$, applying the Cauchy-Schwartz inequality
(i.e. $\left|\sum_ix_i{y}^*_i\right|^2\le\sum_i|x_i|^2 \sum_j|y_j|^2$) to
the sum over $i$ one obtains:
\begin{equation}
\overline{\delta T_{ab}\delta T_{a'b'}}\le
\sum_{m,m'}\sum_i |A^{\{m\}}_i|^2\sum_j |A^{\{m'\}}_j|^2
=\overline{T}_{ab}\overline{T}_{a'b'}.
\end{equation}
This inequality proves that the condition $C_{aba'b'}<1$ is always
fulfilled, even in the case of internal degeneracy. 

To pinpoint the error in~\cite{PRL}, please note that in our
Eq.~\eqref{corr2}, the sum over the internal states is
\textit{outside} the modulus square in contradiction with the 
Eq.(5) of the Letter, where this sum appears \textit{inside} the modulus
square. This results in a completely different mesoscopic situation.
From Eq.~\eqref{corr2}, the intensity correlation arises as an
incoherent sum of the square of different correlation diffusons (one
for each ${\{m\},\{m'\}}$ pair), whereas in Eq.(7) of the Letter, the authors
are calculating the square of a correlation diffuson in which all
${\{m\},\{m'\}}$ contributions add coherently. Obviously, transforming a sum
of intensities into a square of a sum of amplitudes can lead to a very
different physical behavior, like an exponential grow with the size of
the system.


The preceding points, alone, are enough to prove that the large
intensity fluctuations calculated by the Authors are unphysical.
Still, we would like to emphasize an additional crucial point: as
explained by the Authors, the dominant contribution in Eq.~\eqref{corr1}
is obtained only
when the two paths $i$ and $j$ are not sharing any scatterers. In the
case of internal degeneracy, this condition actually implies that,
along these two paths, the atoms can only undergo Rayleigh transitions
(i.e. for which the initial and final internal states are the same),
in full agreement with the experimental results of
Ref.~\cite{rayleigh}. Indeed, the fact that both photon paths $i$
and $j$ correspond to the same global configuration of internal states
(see Eq.\eqref{corr1}), (i.e.  $\{m\}$ during the first pulse and
$\{m'\}$ during the second pulse), means that, if a Raman transition
occurs for a given atom along the photon path $i$, then the same atom
must undergo the same Raman transition along the path $j$: $i$ and $j$
must share, at least, this particular atom. One must mention that
this point is also crucial for the coherent backscattering effect: the
latter arises from interference between photon paths not only
visiting the same atoms in the reverse order, but also along which
each atom performs the same internal transition~\cite{crossedth},
resulting in a lower enhancement factor, in perfect agreement with
the experimental observations~\cite{crossedexp}. In the present case,
this means that only photon paths along
which all the atoms have undergone Rayleigh transitions contribute to
the disorder averaged intensity correlations. Since the number of
those paths is smaller than the total possible paths, we expect
$\overline{T_{ab}T_{a'b'}}< \overline{T}_{ab}\overline{T}_{a'b'}$,
i.e.  $C_{aba'b'}<1$. Only when the ground state is not degenerate
(i.e.  Rayleigh scatterers), all paths contributing to the average
intensity also contribute to the correlation function, leading to
$x_i=y_i$ and the Cauchy-Schwartz inequality becomes an equality, 
implying $C_{aba'b'}=1$.
 
\bigskip 

\begin{small}

\noindent B.~Grémaud${}^1$, D.~Delande${}^1$, C.A.~Müller${}^2$, and C.~Miniatura${}^3$

${}^1$Laboratoire Kastler Brossel

$\phantom{{}^1}$75005 Paris, France

${}^2$Physikalisches Institut, Universität Bayreuth

$\phantom{{}^2}$95440 Bayreuth, Germany

${}^3$Institut Non Linéaire de Nice
 
$\phantom{{}^3}$06560 Valbonne, France
\end{small}

\end{document}